\documentclass[twoside,fleqn]{article}
\usepackage{espcrc2,texdraw,epsfig}
\title{Vortices versus monopoles in color confinement
\thanks{talk presented by A. Di Giacomo}}
\author{L. Del Debbio$^a$, A. Di Giacomo$^a$, B. Lucini$^b$
\vskip 5mm
$^a$ Dipartimento di Fisica, Universit\`a di Pisa,
and INFN Sezione di Pisa, Italy \\
$^b$ Theoretical Physics, University of Oxford, UK}

\begin{document}

\begin{abstract}
We construct the creation operator of a vortex for $SU(2)$ pure gauge
theory using the methods
developed for monopoles. We interpret its vacuum expectation value as
a disorder parameter for the deconfinement phase transition and find
that it behaves in the vacuum in a similar way to monopoles. Results
are extrapolated to the thermodynamical limit using finite-size scaling.
\end{abstract}

\maketitle

\section{Introduction}
Two kinds of topological defects were proposed by 't~Hooft as possible
configurations condensing in the disordered phase of $SU(N)$ Yang-Mills
theories to produce confinement: $Z_N$ vortices~\cite{thooft78}  and
monopoles~\cite{thooft81}.

Monopoles have the natural topology to three dimensional space,
corresponding to a mapping: $S_2 \to SU(2)$. Their condensation
implies dual superconductivity of the vacuum, and gives an appealing
physical picture of confinement in terms of dual Abrikosov tubes
produced by dual Meissner effect~\cite{dualsup}. Indeed, condensation
of magnetic charge has been unambiguously demonstrated by numerical
simulations on the lattice~\cite{adg99}. An operator $\mu$ carrying
non-zero magnetic charge is constructed, and its vev $\langle \mu
\rangle$ is measured in the confined and deconfined phases, as a
candidate disorder parameter. The extrapolation to the thermodynamical
limit is done by finite size scaling techniques and the result is:
\begin{eqnarray}
  && \langle \mu \rangle \neq 0,~~\mathrm{for }~~T<T_c, \\ 
  && \langle \mu \rangle = 0,~~\mathrm{for }~~T>T_c \\ 
  && \langle \mu \rangle \propto \left(1-T/T_c\right)^\delta,
  ~~\mathrm{for }~T \simeq T_c
\end{eqnarray}
Both $T_c$ and the critical index of the correlation length $\nu$ are
obtained, and they agree with determinations by other techniques~\cite{boyd95}. 
The values of $\delta$ are:
\begin{equation}
\begin{array}{l}
\delta = 0.25(10) ~~~\mathrm{for }~~SU(2) \\
\delta = 0.54(4) ~~~\mathrm{for }~~SU(3) \\
\end{array}
\end{equation}
$\langle \mu \rangle \neq 0$ signals dual superconductivity.

The same disorder parameter can be defined in full $QCD$, and shows a similar
behaviour at the deconfinement phase transition~\cite{adg01}. This phenomenon seems to be
independent of the Abelian projection chosen to define the monopoles. The
meaning of these results is that the (yet unknown) excitations, which condense
in the confined phase and are weakly interacting in the dual picture, must be
magnetically charged with respect to all abelian projections. Attempts to
identify these excitations with the monopoles of the maximal abelian
projection do not seem to work as expected~\cite{ambjorn00,polikarpov00}.

In order to understand better the properties of the dual excitations, we
investigate by similar techniques the role of vortices~\cite{ldd00}. In 3+1
dimension vortices are string-like topological defects, associated to closed
curves $C$. The operator $B(C)$ which creates such a vortex obeys the following
algebraic relation with the Wilson loop $W(C^\prime)$, defined as the parallel
transport along the curve $C^\prime$:
\begin{equation}
\label{eq:commut}
B(C) W(C^\prime) = W(C^\prime) B(C) \exp \left( 
\frac{i 2 \pi n_{CC^\prime}}{N} \right)
\end{equation}
with $n_{CC^\prime}$ being the linking number of the curves $C$ and
$C^\prime$~\cite{thooft78}.

It follows from Eq.~\ref{eq:commut} that whenever $\langle W(C^\prime)
\rangle$ obeys the area law, $\langle B(C) \rangle$ obeys the
perimeter law and viceversa~\cite{thooft78}. A consequence of the area
law for $\langle W(C^\prime) \rangle$ is that the Polyakov line,
defined as the parallel transport along a line in the time direction
closed by periodic boundary conditions (pbc), has to vanish.  Instead,
it can be that $\langle P(x) \rangle \neq 0$ when the Wilson loop
obeys the perimeter law. We argue that the same happens for the dual
loops $B(C)$. We define a disorder parameter $\langle \mu \rangle$ as
the ``dual Polyakov line'', corresponding to the operator $B(C)$ for a
curve $C$ going through the lattice, e.g.  parallel to the $z$ axis,
at fixed time and closed by pbc. In the phase where $\langle B(C)
\rangle$ obeys the area law (and therefore $\langle W(C^\prime)
\rangle$ obeys a perimeter law), $\langle \mu \rangle = 0$; whenever
$\langle \mu \rangle \neq 0$ the perimeter law is possible for
$\langle B(C) \rangle$. This observation helps in defining a disorder
parameter in 3+1 dimension.

In 2+1 dimension the vortex is point-like, it can be described by a
local field, and a conserved quantum number can be associated to it,
which is broken in the disordered phase~\cite{thooft78}. Instead, no
conserved quantum number can be associated to vortices in 3+1
dimension.

\section{Creation operator of a vortex}
For the sake of definiteness, let us take for the curve $C$ a rectangle $R$
in the $xy$ plane:
\begin{eqnarray}
  R &=& \left\{ (x,y,z) \right. : \nonumber \\
 &&  \left. x_0<x<x_1, y=y_0, z_0<z<z_1 \right\}
\end{eqnarray}
The definition given below can be extended trivially to any curve $C$.

We define the vacuum expectation value (vev) $\langle B(C) \rangle \equiv \tilde{Z}
/ Z$, where $Z$ is the ordinary partition function for pure Yang-Mills
theories on the lattice defined by the Wilson action:
\begin{equation}
  S[U] = \sum_{x,\mu,\nu} \, \mathrm{Re }~\mathrm{Tr } \left[ 1 - P_{\mu\nu}
  \right] 
\end{equation}
with $P_{\mu\nu}$ is the usual plaquette. $\tilde Z$ is the partition
function corresponding to the action $\tilde S$ obtained from $S$ by the
change:
\[
  \begin{array}{l}
    P_{0y}(t_0,x_0<x<x_1, y=y_0, z_0<z<z_1) \mapsto \\
    e^{i 2 \pi/N} P_{0y}(t_0,x_0<x<x_1, y=y_0, z_0<z<z_1)
  \end{array}
\]
We study by numerical simulations, the be\-ha\-viour of the "dual Polyakov line",
which corresponds to the above definition with:
\begin{eqnarray}
  z_0 &\to& -\infty \nonumber \\
  z_1 &\to& +\infty \nonumber 
\end{eqnarray}
in that case the change of variables becomes: 
\begin{equation}
  \label{eq:def}
  \begin{array}{l}
    P_{0y}(t_0,x_0<x<x_1, y=y_0, z) \mapsto  \\
    e^{i 2 \pi/N} P_{0y}(t_0,x_0<x<x_1, y=y_0, z), \\
    ~~~\mathrm{for~all}~~z
  \end{array}
\end{equation}
For the proof that Eq.~\ref{eq:def} really corresponds to the definition of
$B(C)$ as defined by Eq.~\ref{eq:commut} and for comparison with alternative
definitions in the literature, we refer to~\cite{ldd00}.

We shall also compute the correlator:
\begin{equation}
\label{eq:corr}
\Gamma(t) = \langle \mu(t_0,x_0,y_0) \,\mu(t_0+t,x_0,y_0) \rangle
\end{equation}
As $t\to\infty$,  
\[
\Gamma(t) \sim A e^{-m t} + \langle \mu \rangle^2
\]
whence $\langle \mu \rangle$ can be extracted.

At finite $T$ there is no propagation in time and $\langle \mu \rangle$ is
computed directly, provided $C^*$ boundary conditions are adopted~\cite{ldd00,adg99}.

\section{Numerical results}
\begin{figure}[htp]
\begin{center}
\epsfig{figure=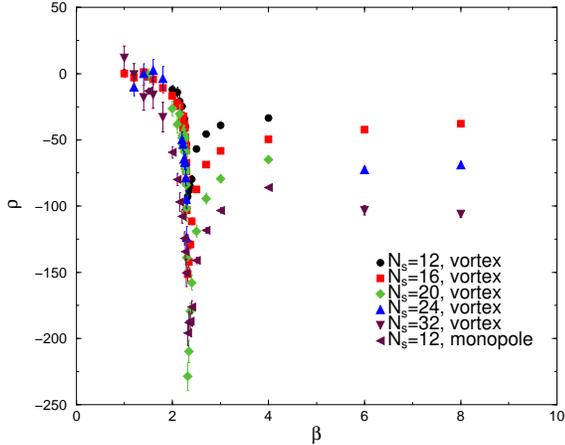, angle=270, width=7.5cm} 
\end{center}
\caption{Comparison of $\rho$ defined with the vortex
and the monopole creation operator.} 
\label{fig:peak}
\end{figure}

We compute $\langle \mu \rangle$ by the techniques explained above, or better
we compute:
\begin{equation}
\rho = \frac{\mbox{d}}{\mbox{d} \beta} \log \langle \mu \left( t_0, x_0, y_0
\right) \rangle = \langle S \rangle _S - \langle \tilde{S} \rangle _{\tilde{S}}
\end{equation}
which contains the same information and is less
noisy~\cite{ldd95,adg97,adg99}. The behaviour of $\rho$ vs $\beta$ for a $N_t
\times N_s^3$ lattice with $N_t=4$ and $N_s=12,16,20,24,32$ is plotted in
Fig.~\ref{fig:peak}; the corresponding $\rho$ for the creation of monopoles is
plotted in the same figure for comparison. As explained in the introduction,
$\rho$ is independent within errors of the Abelian projection
used to define the magnetic charges~\cite{adg99}. 

Similarly to the case of monopoles, the behaviour of $\rho$ at large
$\beta$ is proportional to $N_S$
\[
\rho \simeq -4 N_s +C
\]
implying that $\langle \mu \rangle = \exp\left[\int_0^\beta \,d\beta^\prime
\, \rho(\beta^\prime) \right]$ vanishes for $\beta>\beta_c$ as $N_s \to
\infty$. At low $\beta$, $\rho$ is bounded from below when the volume is
increased, which implies $\langle \mu \rangle \neq 0$ for $\beta < \beta_c$. 

Around $\beta_c$, $\langle \mu \rangle \sim (\beta_c-\beta)^\delta$. In the
scaling region one expects:
\[
\langle \mu \rangle \sim (\beta_c-\beta)^\delta \Phi(\frac{N_s}{\xi})
\]
where $\xi \sim (\beta_c-\beta)^{-\nu}$ is the correlation length.

\begin{figure}[htp]
\begin{center}
\epsfig{figure=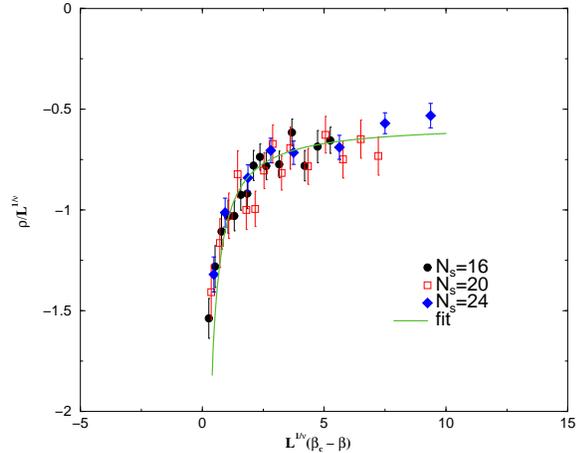, angle=270, width=7.5cm} 
\end{center}
\caption{Plot of rescaled data.} 
\label{fig:scaling}
\end{figure}

This yields the scaling law:
\begin{equation}
\frac{\rho}{L^{1/\nu}} = f\left(L^{1/\nu}
\left(\beta_c - \beta\right)\right)
\end{equation}
We shall assume $f(x)=-\delta/x + c$. Requiring scaling fixes $\beta_c,
\nu$ and $\delta$. The quality of the scaling for $SU(2)$ is shown in
Fig.~\ref{fig:scaling}. 

The critical exponents that we obtain are:
\[
\beta_c=2.30(1), \delta=0.5(1), \nu=0.7(1)
\]
$\beta_c$ and $\nu$ are compatible with the standard
determinations~\cite{boyd95}. $\delta$ is equal within errors to the corresponding
index for the monopole disorder parameter. Preliminary data for $SU(3)$
present a similar behaviour~\cite{ldd00b}. 

We conclude that the ``dual Polyakov line'' is a good disorder
parameter for confinement. This is a confirmation of the argument in
Ref.~\cite{thooft78} about the area and perimeter law. A direct check
is presented in Ref.~\cite{delia00}.

\section{Conclusions and outlook}
A disorder parameter for confinement, related to the condensation of
magnetic charges already exists~\cite{ldd95,adg97}, its behaviour is
independent of the abelian projection used to define the monopoles,
and describes confinement both in quenched~\cite{adg99} and unquenched
QCD~\cite{adg01}. This means that dual superconductivity is at work as
a mechanism of confinement. Assuming that there exists a dual
description in terms of weakly coupled fields, they must carry
magnetic charge with respect to any abelian projection.

The result of the present investigation on vortices has no direct
implication on the symmetry of the dual fields. It confirms, however,
the argument of Ref.~\cite{thooft78} that $\langle B(C) \rangle$ obeys
the area law when $\langle W(C^\prime) \rangle$ obeys the perimeter
law and viceversa, which has also been tested in~\cite{delia00}. The
dual Polyakov line is a good disorder parameter for confinement, in
the same way as the ordinary Polyakov line is a good order
parameter. An analogous analysis for $SU(3)$ is in preparation and
gives similar results~\cite{ldd00b}. A measurment of the critical indices using a
monopole operator in full $QCD$ is in progress~\cite{adg01}. We are
also trying to extend to full QCD the study of the dual Polyakov
line. The dependence on $N_c$ of both approaches is an interesting
question, and deserves numerical investigation, in order to check the
basic ideas of the large-$N_c$ limit.

\noindent
{\bf Acknowledgements} We thank M. D'Elia, G. Paffuti, M. Teper and A. Kovner
for enlightening discussions. Financial support by the EC TMR Pro\-gram
ERB\-FMRX-CT97-0122 and by MURST is acknowledged. BL is funded by PPARC under
Grant PPA/G/0/1998/00567.

\end{document}